\documentclass[journal]{IEEEtran}
\usepackage{amsmath,amssymb,amsfonts}
\usepackage{algorithm}
\usepackage{algorithmic}
\usepackage{graphicx}
\usepackage{textcomp}
\usepackage{xcolor}
\definecolor{myred}{RGB}{205 38 38}
\usepackage{xcolor}
\def\BibTeX{{\rm B\kern-.05em{\sc i\kern-.025em b}\kern-.08em
    T\kern-.1667em\lower.7ex\hbox{E}\kern-.125emX}}
\usepackage{amsthm}
\usepackage{bm}

\usepackage{multirow}
\usepackage{array}
\usepackage{booktabs}
\usepackage{lipsum}
\usepackage{xcolor}
\usepackage{etoolbox}
\usepackage{algorithm}
\usepackage{algorithmic}
\usepackage{amsthm}

\usepackage{lineno,hyperref}
\modulolinenumbers[5]
\usepackage{bm}
\usepackage{multirow}
\usepackage{array}
\usepackage{booktabs}
\usepackage{lipsum}
\usepackage{xcolor}
\usepackage{etoolbox}
\usepackage{cite}
\ifCLASSINFOpdf
\else
\fi
\ifCLASSOPTIONcompsoc
  \usepackage[caption=false,font=normalsize,labelfont=sf,textfont=sf]{subfig}
\else
  \usepackage[caption=false,font=footnotesize]{subfig}
\fi

\newcommand{\xref}{\mathbf{x}_{\rm ref}}
\newcommand{\bd}{\mathbf}

\graphicspath{{./FiguresJournal/}} 

\begin{document}

\title{Inverter Control with Time-Varying and Nonconvex State and Input Constraints
}
\author{Zixiao~Ma,~\IEEEmembership{Member,~IEEE,}
	Baosen~Zhang,~\IEEEmembership{Member,~IEEE,}

 \thanks{(\emph{Corresponding author: Zixiao Ma})}
	\thanks{Z. Ma is with the Department of Electrical and Computer Engineering, SUNY Binghamton University, Binghamton, NY, 13902, USA (E-mail: zma10@binghamton.edu}
	\thanks{B. Zhang is with the Department of Electrical and Computer Engineering, University of Washington, Seattle, WA, 98195, USA (e-mails: zhangbao@uw.edu).}
    \thanks{The authors were partially supported by the Clean Energy Institute at the University of Washington.}}
\maketitle

\begin{abstract}
The growing integration of inverter-based resources (IBRs) into modern power systems poses significant challenges for maintaining reliable operation under dynamic and constrained conditions. This paper focuses on the power tracking problem for grid-connected IBRs, addressing the complexities introduced by voltage and power factor constraints. Voltage constraints, being time-varying and nonlinear input constraints, often conflict with power factor constraints, which are state constraints. These conflicts, coupled with stability requirements, add substantial complexity to control design. To overcome these challenges, we propose a computationally efficient static state-feedback controller that guarantees stability and satisfies operational constraints. The concept of achievability is introduced to evaluate whether power setpoints can be accurately tracked while adhering to all constraints. Using a parameterization framework and the S-lemma, we develop criteria to assess and maximize the continuous achievable region for IBR operation. This framework allows system operators to ensure safety and stability by precomputing a finite set of control gains, significantly reducing online computational requirements. The proposed approach is validated through simulations, demonstrating its effectiveness in handling time-varying grid disturbances and achieving reliable control performance.
\end{abstract}


\section{Introduction}
The increasing integration of inverter-based resources (IBRs) into modern power grids offers significant opportunities for the use of renewable energy but also introduces new challenges for the operation and control of the grid. A standard requirement in IBR operations is power tracking, where the goal is to design controllers that regulate active and reactive powers to desired setpoints while satisfying operational constraints. This paper focuses on controller designs that ensure fast convergence to power setpoints while meeting the state, input, and stability constraints inherent in IBR operation \cite{Chen2009,Zhang2021,Ma2021}.

The constraints that arise in IBR power tracking are nontrivial. In this paper, we consider two of them in detail: voltage and power factor constraints. Voltage constraints capture the requirment that the output voltage of the inverter remain within a specified range around the nominal operating point to ensure safe operation. Unlike the linear and constant voltage constraints seen in conventional machines, voltage constraints in IBRs are nonlinear and time-varying, depending on the grid voltage and control inputs \cite{Ma2023,Ma2023b,Du2019,Ma2024}. The power factor constraint, which ensures the quality of power delivered by the IBR \cite{Zhang2023}.

However, managing both the voltage and the power factor constraints simultaneously in direct power control (DPC) of inverters poses significant challenges. In the formulation of this paper, the power factor constraint is a state constraint, while the voltage constraint is an input constraint. These constraints can conflict, limiting the feasible operating range of the IBR. When combined with the standard stability requirements, simultaneously satisfying these constraints becomes even more complex \cite{Sun2023}.

Model predictive control (MPC) has been widely adopted for its ability to explicitly handle system constraints, including control input limitations, in DPC applications \cite{Hu2015,Xiao2024}. However, MPC faces limitations due to its high computational requirement and the possibility of running into infeasible problems, particularly under time-varying grid voltage conditions. Robust or adaptive versions have been proposed, but they lead to more complex computational problems that are difficult to implement~\cite{Zhang2020}. 

There are also a host of simpler DPC controllers~\cite{Noguchi1998,Hu2011,Li2017,Gui2018,fu2021hybrid,Gui2023}. Early methods, such as lookup table-based DPC, directly select switching states based on real-time measurements like power errors and terminal voltage \cite{Noguchi1998}. However, these methods often result in variable switching frequencies, leading to harmonic distortion and challenges in filter design. Sliding mode control and grid-voltage-modulated DPC improve robustness and response time but suffer from issues like chattering and lack of guaranteed convergence \cite{Hu2011,Li2017}. Advanced approaches, such as port-controlled Hamiltonian DPC, introduce dissipative properties to reduce power ripples but do not fully address operational constraints \cite{Gui2023}.

Despite progress in DPC controller design, the need to simultanenously satisfy multiple constraints remains challenging. Some methods adopt conservative approximations of the constraint set, but their applicability to timevarying and nonlinear systems like IBRs are limited~\cite{Kothare1996,Mayne2000,Chisci2001}. Others assume constraints are inherently satisfied, which may not hold under significant changes in power setpoints or grid disturbances. To address this, the concept of achievability was introduced in our earlier work \cite{Ma2024control}. A power setpoint is deemed achievable if there exists a control action that ensures accurate power tracking while satisfying all constraints. 

This paper proposes a novel control strategy for grid-connected inverter-based resources (IBRs) that bridges the gaps in computational efficiency, simultaneous satisfaction of constraints, and resilience to time-varying grid disturbances. Specially, by parameterizing constraints in terms of the control gain and power setpoints, we use the S-lemma to derive achievability criteria and efficiently evaluate feasible operating regions. The proposed framework ensures reliable and efficient operation of grid-connected IBRs under realistic conditions. The main contributions are summarized as follows:

\begin{itemize}
\item We develop an efficient static state-feedback controller that ensures stability and satisfies state and input constraints simultaneously. It is much easier to implement than online optimization approaches like MPC.

\item A systematic approach is introduced to evaluate the achievability of power setpoints. This framework guarantees that, under bounded arbitrary and time-varying grid voltages, the generated control signals respect constraints on the IBR output voltage and power factor.

\item We propose a methodology to determine and maximize the achievable region. This allows system operators to ensure stability and safety by storing a finite set of precomputed control gains and switching between them as needed.
\end{itemize}

The remainder of the paper is structured as follows. Section \ref{C2} discusses the DPC problem for grid-connected IBRs under time-varying constraints. Section \ref{C3} presents the proposed static state-feedback controller and the parameterization of constraints in terms of control gain and power setpoints. Section \ref{C4} develops the optimization model for the power controller and achievability evaluation. Section \ref{C5} provides simulation results validating the proposed approach, and Section \ref{C6} concludes the paper.

\section{Problem Statement}\label{C2}
\begin{figure}[t!]
		\centering
		\includegraphics[width=0.95\columnwidth]{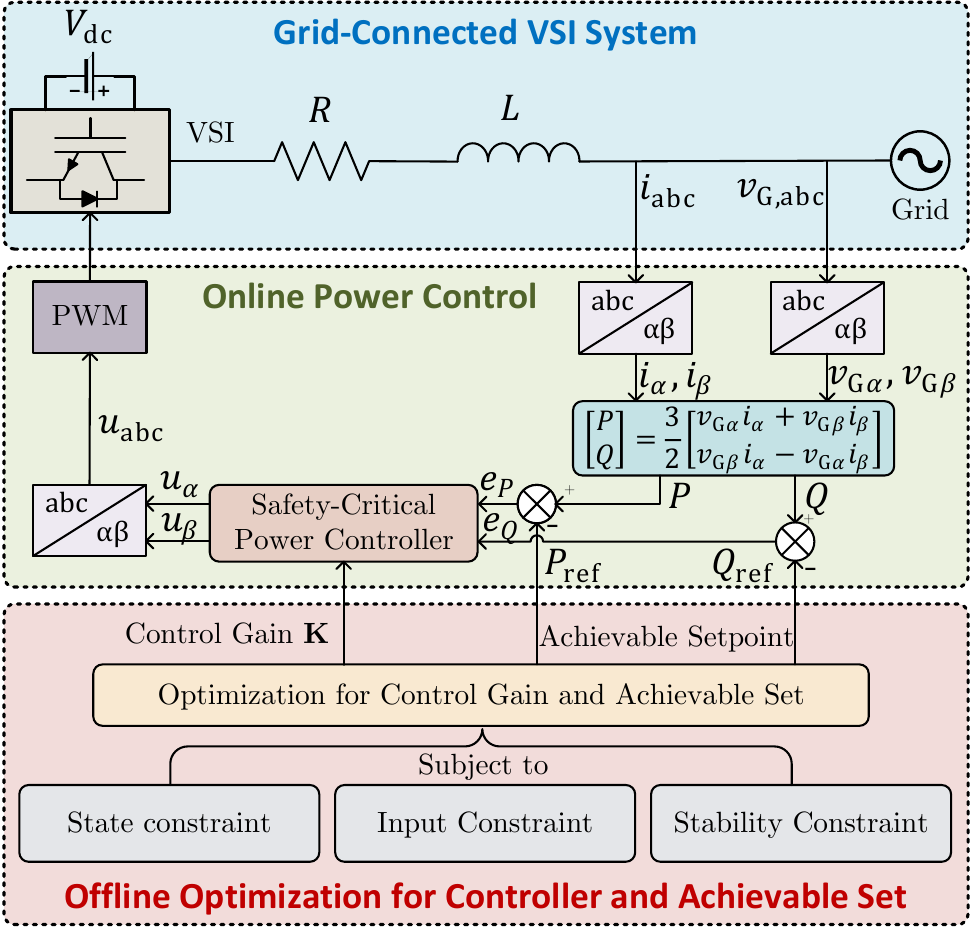}
		\caption{Overview of the proposed power control scheme for a grid-connected IBR with state, input, and stability constraints. The online safety-critical power controller is implemented as a static state-feedback controller to ensure computational efficiency. The control gain and achievable power setpoints are determined through an offline constrained optimization process, effectively shifting the computational burden from the online stage to the offline stage.}
		\label{systemfig}
	\end{figure}

In this section, we present the model of a  grid-connected IBR. In particular, we consider a large-signal framework and then linearize the system through the introduction of an auxiliary controller. State constraints are then defined, including  constraints on the power factor and voltage magnitude. The voltage constraint is then transformed into a time-varying, non-convex constraint on the auxiliary control input. Finally, the power control problem for the grid-connected inverter is formulated. An overview of the problem framework is provided in Fig. \ref{systemfig}.
\subsection{Model of Grid-Connected IBR}
We aim to control the IBR output power and determine the achievability of power setpoints of an IBR. Thus, we adopt the DPC model in \cite{Gui2018,xia2022existence,ayalew2021enhanced}: 
\begin{subequations}\label{system_pq}
   \begin{align}
    \dot{P}&=-\frac{R}{L}P-\omega Q+\frac{3}{2L}\left(v_{{\rm G}\alpha}(t)u_{\alpha}+v_{{\rm G}\beta}(t)u_{\beta}-V_{\rm G}(t)^2\right),\\
    \dot{Q}&=\omega P-\frac{R}{L} Q+\frac{3}{2L}\left(v_{{\rm G}\beta}(t)u_{\alpha}-v_{{\rm G}\alpha}(t)u_{\beta}\right),\\
        & v_{{\rm G}\alpha}(t)=V_{\rm G}(t)\cos{\omega t},\\
   &  v_{{\rm G}\beta}(t)=V_{\rm G}(t)\sin{\omega t},
\end{align} 
\end{subequations}
where $P$ and $Q$ are active and reactive power, respectively. $R$ and $L$ are the resistance and inductance of the filter, respectively. $\omega=2\pi f$ is the angular frequency and $f$ is the grid frequency. $V_{\rm G}(t)$ is the grid voltage, which could be time-varying. For notational simplicity, we often suppress the time index and write it as $V_{\rm G}$.  The control inputs are $u_\alpha$ and $u_\beta$, the inverter output voltages in $\alpha-\beta$ coordinates.

The system in~\eqref{system_pq} is nonlinear because the controls $u_\alpha$ and $u_\beta$ enter through multiplications with time-varying grid voltages. Therefore, we introduce an auxiliary controller:
\begin{align}\label{def_input}
   \mathbf{u}=\begin{bmatrix}
        u_P\\
        u_Q
    \end{bmatrix}=\begin{bmatrix}
        v_{{\rm G}\alpha}u_{\alpha}+v_{{\rm G}\beta}u_{\beta}\\
        v_{{\rm G}\beta}u_{\alpha}-v_{{\rm G}\alpha}u_{\beta}
    \end{bmatrix}.
\end{align}
Defining state vector as $\mathbf{x}=[P\;\;Q]^\top$, the direct power control mode of grid-connected IBR (\ref{system_pq}) becomes:
\begin{align}\label{system}
\dot{\mathbf{x}}=\mathbf{A}\mathbf{x}+\mathbf{B}\mathbf{u}+\mathbf{E}V_{\rm G}^2.
\end{align}
where the square of the grid voltage magnitude $V_{\rm G}^2$, acts as a time-varying disturbance, and the system matrices are 
\begin{align*}
    \mathbf{A}=\begin{bmatrix}
       -\frac{R}{L} & -\omega\\\omega&-\frac{R}{L}
    \end{bmatrix}, \;\mathbf{B}=\begin{bmatrix}
        \frac{3}{2L}&0\\0& \frac{3}{2L}
    \end{bmatrix},\;\;\mathbf{E}=\begin{bmatrix}
        -\frac{3}{2L}\\ 0
    \end{bmatrix}.
\end{align*}

It is easy to recover the original inverter control signals $u_\alpha$ and $u_\beta$ from the auxiliary conotroller $\mathbf{u}$ through:
\begin{align}
    \begin{bmatrix}
        u_{\alpha}\\u_{\beta}
    \end{bmatrix}&=\begin{bmatrix}
          v_{{\rm G}\alpha}&v_{{\rm G}\beta}\\
        v_{{\rm G}\beta}&-v_{{\rm G}\alpha}
    \end{bmatrix}^{-1}\begin{bmatrix}
        u_P\\
        u_Q
    \end{bmatrix}\\
    &=\frac{1}{V_{\rm G}}\begin{bmatrix}
        \cos{\omega t}&\sin{\omega t}\\
        \sin{\omega t}&-\cos{\omega t}
    \end{bmatrix}\mathbf{u}.
\end{align}

\subsection{Designing State and Input Constraints for Power and Voltage Safety in IBR}
\begin{figure}[ht!]
		\centering
		\includegraphics[width=0.65\columnwidth]{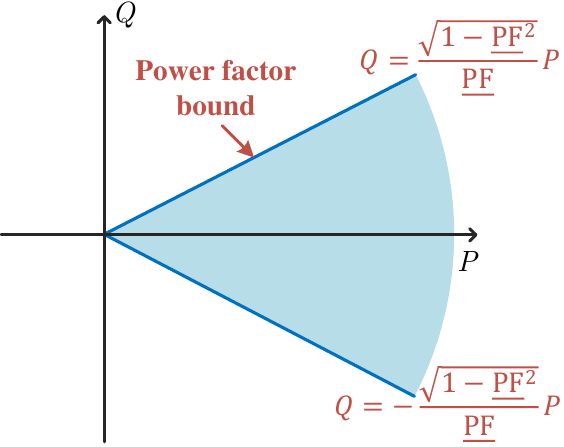}
		\caption{The blue region denotes the state constraint on the power factor. The slope of the sector can be calculated from (\ref{stateconstraint}). }
		\label{stateconstraintfig}
	\end{figure}

There are two commonly imposed constraints on the system. The first is a constraint on the power factor at the output of the inverter.  To guarantee the power quality, as shown in Fig. \ref{stateconstraintfig}, the power factor should be bounded within the desired range during the power control of the inverter, i.e.,
\begin{align}\label{pfconstraint}
    \frac{P}{\sqrt{P^2+Q^2}}\geqslant \underline{\rm PF},
\end{align}
where $\underline{\rm PF}$ denotes the lower bound of power factor, and usually $\underline{\rm PF}=0.95$. To see this is a state constraint, note that (\ref{pfconstraint}) can be written as:
\begin{align}\label{stateconstraint}
    (1-\underline{\rm PF}^2)x_1^2-\underline{\rm PF}^2x_2^2\geqslant0.
\end{align}


The second constraint is on the voltage of the IBR. To remain operational, the output voltage, $U$, of the IBR should remain within a prescribed bound even afer a disturbance. Denoting $\underline{U}$ and $\overline{U}$ as the lower and upper bounds of the IBR output voltage magnitude $U$, respectively, $U$ should satisfy 
\begin{align}
    \underline{U}\leqslant U\leqslant\overline{U}.
\end{align}
Based on the definition of auxiliary controller \ref{def_input}, we have $\|\mathbf{u}\|_2=\sqrt{u_P^2+u_Q^2}=UV_{\rm G}$. Thus, the constraint on the control input of the IBR system (\ref{system}) is developed as
\begin{align}\label{inputconstraint}
   \underline{U}V_{\rm G}\leqslant \|\mathbf{u}\|_2\leqslant \overline{U}V_{\rm G}.
\end{align}

Here we make an assumption about $V_{\rm G}$. We assume that while it is time-varying, it is bounded within the range $[\underline{V_{\rm G}},\overline{V_{\rm G}}]$. Therefore, it yields a disturbance constraint as follows,
\begin{align}\label{dconstraint}
    \underline{V_{\rm G}}^2\leqslant V_{\rm G}^2\leqslant\overline{V_{\rm G}}^2.
\end{align}
This assumption is not overly restrictive, since it mainly ensures that $V_{\rm G}$ cannot be too large (the grid voltage cannot be very high). 

\begin{figure}[ht!]
		\centering
		\includegraphics[width=0.95\columnwidth]{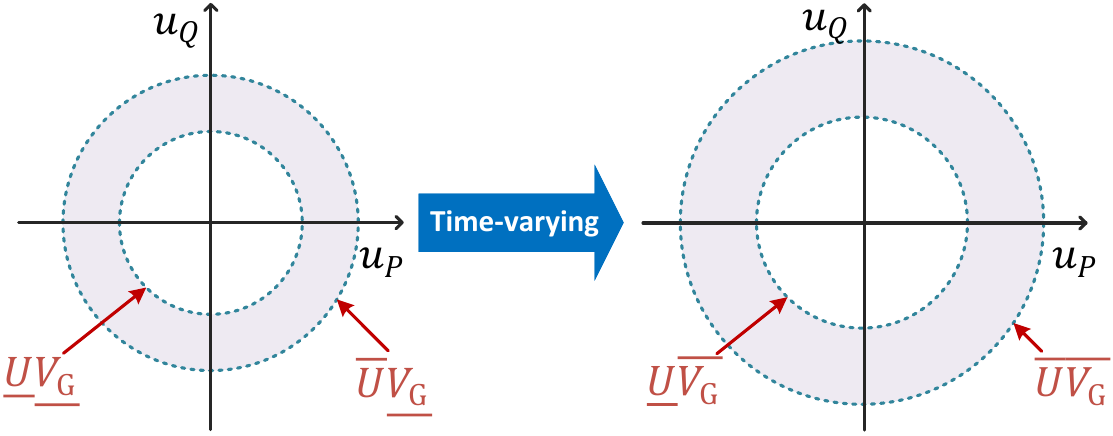}
		\caption{The circle regions represent the nonconvex input constraints. The achievable control signal depends on the time-varying value of $V_{\rm G}$. }
		\label{inputconstraintfig}
	\end{figure}

{\emph{Remark 1:} As illustrated in Fig. \ref{inputconstraintfig}, the input constraint (\ref{inputconstraint}) is both nonconvex and time-varying, a characteristic that is somewhat unique to inverter control. Simply removing the lower bound to convexify feasible region can remove its low voltage ridethrough capability, a key requirement for IBRs~\cite{yang2013low}.  Incorporating the additional state constraint further complicates the controller design, as it requires simultaneously satisfying both state and input constraints. Existing control methods are generally unsuitable for addressing such a complex problem. While iterative optimization-based approaches, such as MPC, can provide a solution, they impose a significant computational burden for determining each control signal~\cite{ademola2021mpc}. This can lead to violations of the computational speed requirements. We will demonstrate these challenges in the simulation section. }

\subsection{Power Control Problem Statement}
The control objective, as illustrated in Fig. \ref{systemfig}, is to regulate the IBR output active and reactive powers to their designated setpoints while ensuring stability and satisfying the state constraint (\ref{stateconstraint}) and the time-varying nonconvex input constraint (\ref{inputconstraint}). These constraints limit the operation of the grid-connected IBR, as not all setpoints are achievable with a single universal controller. A setpoint is said to be achievable if there exists a static feedback controller $\mathbf{u}$  that satisfies both constraints and ensures $\mathbf{x}(t) \to \mathbf{x}_{\rm ref}$ with exponential convergence. The collection of all such achievable setpoints for a given control input is referred to as the achievable set, with formal definitions provided in Section~\ref{C3}.

This paper addresses three key questions. First, how can a computationally efficient controller be designed to achieve the stated objective? Second, what is the achievable set for such a control algorithm? Third, how can the controller be designed to maximize this achievable set? These questions guide the development presented in the remainder of this paper.

\section{Parameterization of State and Input Constraints for Safe Power Controller Design}\label{C3}
In this section, a static state-feedback power controller is designed to facilitate real-time control of the inverter. The stability, state, and input constraints are subsequently parameterized with respect to the static control gain and power setpoints using Lyapunov theory and the S-lemma. These parameterized constraints will be used as optimization constraints in the controller synthesis presented in the next section.
\subsection{Power Controller Design}
The controllability of system (\ref{system}) and the assumption that the time-varying disturbance $V_{\rm G}^2$ is measurable allow the use of a straightforward feedforward controller to regulate the output power to the desired setpoints~\cite{Gui2018,Gui2019}: $\mathbf{u}=-\mathbf{B}^{-1}\mathbf{A}\mathbf{x}_{\rm ref}-\mathbf{B}^{-1}\mathbf{E}V_{\rm G}^2.$ Compared to MPC-based controllers, this open loop controller is  easier to implement because of its low computational requirements and is thus suited to real-time applications. The drawback of this controller is that it does not have degrees of freedom for optimizing the achievable set or control performance. 

To provide flexiblity, we add a static state feedback term as follows:
\begin{align}\label{controller}
    \mathbf{u}=-\mathbf{K}(\mathbf{x}-\mathbf{x}_{\rm ref})-\mathbf{B}^{-1}\mathbf{A}\mathbf{x}_{\rm ref}-\mathbf{B}^{-1}\mathbf{E}V_{\rm G}^2,
\end{align}
where $\mathbf{K} \in \mathbb{R}^{2\times 2}$ is feedback gain to be optimized to maximize the size of the achievable set. Defining error state as $\tilde{\mathbf{x}}=\mathbf{x}-\mathbf{x}_{\rm ref}$, the closed-loop error dynamics can be obtained by substituting (\ref{controller}) into (\ref{system}):
\begin{align}\label{errordynamics}
    \dot{\tilde{\mathbf{x}}}=\left(\mathbf{A}-\mathbf{B}\mathbf{K}\right)\tilde{\mathbf{x}}.
\end{align}

We want the control gain $\mathbf{K}$ to satisfy three conditions: 1) stability or the eigenvalues of $\mathbf{A}-\mathbf{B}\mathbf{K}$ have negative real parts; 2) the input constraint in \eqref{inputconstraint}; and, 3) the state constaint in \eqref{stateconstraint}. Each of these conditions has been studied, but there is no existing work that provides a systematic approach to parameterizing the state, input, and stability constraints in conjunction with the linear feedback control gain. In the subsequent subsections, we will employ the S-Lemma to derive conditions that ensure the satisfaction of these constraints. The developed constraints are explicitly dependent on the control gain $\mathbf{K}$ and the setpoints $\mathbf{x}_{\rm ref}$, and therefore can be used to facilitate the optimization of the controller and the achievability analysis in Section \ref{C4}.


\subsection{State Constraints}
We aim to guarantee that the state constraint (\ref{stateconstraint}) is always satisfied with the proposed state feedback controller (\ref{controller}). To this end, let us first define a power safety region as 
\begin{align}\label{SCregion}
    \Omega_{\rm SC}=\{\mathbf{x}|\mathbf{x}^\top \mathbf{Q}_a^{\rm SC}\mathbf{x}+\mathbf{r}_a^{{\rm SC}^\top}\mathbf{x}+{c}_a^{\rm SC}\geqslant 0\}
\end{align}
It is clear that any $\mathbf{x} \in \Omega_{\rm SC}$ satisfies the state constraint in \eqref{stateconstraint}. The following theorem establishes conditions under which $\Omega_{\rm SC}$ is forward-invariant, ensuring that trajectories starting in the set remain within it.

\emph{Theorem 1:} Given a power setpoint $\mathbf{x}_{\rm ref}$, if we can determine a control feedback gain $\mathbf{K}$, such that
\begin{align}\label{theorem1}
    \mathcal{Q}_b^{\rm SC}-\lambda^{\rm SC}\mathcal{Q}_a^{\rm SC}\geqslant0
\end{align}
for some nonnegative constant $\lambda^{\rm SC}$, then, the power safety region (\ref{SCregion}) is controlled to be a forward-invariant set by (\ref{controller}), where
\begin{subequations}\label{SCdefs}
  \begin{align}
  \mathcal{Q}_a^{SC}&=\mathbf{x}^\top \mathbf{Q}_a^{\rm SC}\mathbf{x}+\mathbf{r}_a^{{\rm SC}^\top}\mathbf{x}+{c}_a^{\rm SC},\\
\mathcal{Q}_b^{\rm SC}&=\mathbf{x}^\top \mathbf{Q}_b^{\rm SC}\mathbf{x}+\mathbf{r}_b^{{\rm SC}^\top}\mathbf{x}+{c}_b^{\rm SC},\label{SCdefs_a}\\
   \mathbf{Q}_b^{\rm SC}&=\frac{1}{2}\left(\Tilde{\mathbf{Q}}_b^{\rm SC}+\Tilde{\mathbf{Q}}_b^{{\rm SC}^\top}\right),\label{SCdefs_b}\\
   \Tilde{\mathbf{Q}}_b^{\rm SC}&=\begin{bmatrix}
    2(1-\underline{\rm PF}^2) &0\\0&-2\underline{\rm PF}^2
    \end{bmatrix}(\mathbf{A}-\mathbf{B}\mathbf{K}),\label{SCdefs_c}\\
       \mathbf{r}_b^{\rm SC}&=-\begin{bmatrix}
    2(1-\underline{\rm PF}^2) &0\\0&-2\underline{\rm PF}^2
    \end{bmatrix}(\mathbf{A}-\mathbf{B}\mathbf{K})\mathbf{x}_{\rm ref},\label{SCdefs_d}\\
    c_b^{\rm SC}&=0.\label{SCdefs_e}
\end{align}  
\end{subequations}



\emph{Proof:} We first apply the Invariance Principle to determine the conditions under which $\Omega_{\rm SC}$ is forward-invariant. Then, the S-lemma is used to parameterize these conditions into the form of (\ref{theorem1}).

When the power safety constraint (\ref{stateconstraint}) is satisfied, it has $\mathcal{Q}_a^{\rm SC}\geqslant 0$ for some system state $\mathbf{x}$. We can consider $\mathcal{Q}_a^{\rm SC}$ as a potential function, whose derivative equals
\begin{align}
    \mathcal{Q}_b^{\rm SC}=\;&\dot{\mathcal{Q}}_a^{\rm SC}=\dot{\mathbf{x}}^\top \mathbf{Q}_a^{\rm SC}\mathbf{x}+\mathbf{x}^\top \mathbf{Q}_a^{\rm SC}\dot{\mathbf{x}}+\mathbf{r}_a^{{\rm SC}^\top}\dot{\mathbf{x}}\nonumber\\
    =\;&\begin{bmatrix}
        \dot{x}_1&\dot{x}_2
    \end{bmatrix}\mathbf{Q}_a^{\rm SC}\begin{bmatrix}
        x_1\\x_2
    \end{bmatrix}+\begin{bmatrix}
        x_1&x_2
    \end{bmatrix}\mathbf{Q}_a^{\rm SC}\begin{bmatrix}
        \dot{x}_1\\\dot{x}_2\end{bmatrix}\nonumber\\
        =\;&2(1-\underline{\rm PF}^2)\dot{x}_1x_1-2\underline{\rm PF}^2\dot{x}_2x_2\nonumber\\
        =\;&\mathbf{x}^\top\begin{bmatrix}
            2(1-\underline{\rm PF}^2)&0\\0&-2\underline{\rm PF}^2
        \end{bmatrix}(\mathbf{A}\mathbf{x}+\mathbf{B}\mathbf{u}+\mathbf{E}V_{\rm G}^2)\nonumber\\
        =\;&\mathbf{x}^\top\underbrace{\begin{bmatrix}
            2(1-\underline{\rm PF}^2)&0\\0&-2\underline{\rm PF}^2
        \end{bmatrix}\left(\mathbf{A}-\mathbf{B}\mathbf{K}\right)}_{\Tilde{\mathbf{Q}}_b^{\rm SC}}\mathbf{x}\nonumber\\
        \;&\underbrace{-\mathbf{x}_{\rm ref}^\top\left(\mathbf{A}-\mathbf{B}\mathbf{K}\right)^\top\begin{bmatrix}
            2(1-\underline{\rm PF}^2)&0\\0&-2\underline{\rm PF}^2
        \end{bmatrix}}_{\mathbf{r}_b^{{\rm SC}^\top}}\mathbf{x}.\nonumber
\end{align}

Since the first term of $\mathcal{Q}_b^{\rm SC}$ is a scalar, we have
\begin{align}
    \mathbf{x}^\top \Tilde{\mathbf{Q}}_b^{\rm SC}\mathbf{x}=\mathbf{x}^\top \underbrace{\frac{1}{2}\left(\Tilde{\mathbf{Q}}_b^{\rm SC}+\Tilde{\mathbf{Q}}_b^{{\rm SC}^\top}\right)}_{\mathbf{Q}_b^{\rm SC}}\mathbf{x}.
\end{align}

In this way, we can obtain a symmetric matrix $\mathbf{Q}_b^{\rm SC}$, which is a necessary condition to facilitate the S-Lemma. 

According to the Invariance Principle, if
\begin{align}\label{SCIP}
    \mathcal{Q}_a^{\rm SC}\geqslant0\Rightarrow\dot{\mathcal{Q}}_a^{\rm SC}\geqslant0,
\end{align}
then once $\mathbf{x}$ enters $\Omega_{\rm SC}$, it can never escape from it under control of (\ref{controller}). This implication relationship can be parameterized using the S-Lemma \cite{Polik2007}, i.e., (\ref{SCIP}) holds if and only if there exists a $\lambda_{\rm SC}\geqslant0$ such that
\begin{align*}
    \mathcal{Q}_b^{\rm SC}-\lambda^{\rm SC}\mathcal{Q}_a^{\rm SC}\geqslant0
\end{align*}
given a power setpoint $\mathbf{x}_{\rm ref}$ and feedback control gain $\mathbf{K}$.\hfill$\Box$.

\emph{Remark 2:} Note that inequality (\ref{theorem1}) in \emph{Theorem 1} parameterizes the state constraint (\ref{stateconstraint}) in terms of the control gain $\mathbf{K}$ and the reference point $\mathbf{x}_{\rm ref}$, such that the term $\mathcal{Q}_b^{\rm SC}$ is a function of both $\mathbf{K}$ and $\mathbf{x}_{\rm ref}$. This is particularly valuable because, given a controller with gain $\mathbf{K}$ as defined in (\ref{controller}), \emph{Theorem 1} specifies the set of power setpoints that can be achieved without violating the state constraint (\ref{stateconstraint}) once the trajectory $\mathbf{x}(t)$ enters the region $\Omega_{\rm SC}$.

To eliminate the state variables in (\ref{theorem1}), we use the Schur Complement method \cite{Zhang2006} to convert it into the following Linear Matrix Inequalities (LMI):

\begin{equation}\label{theorem1_lmi}
    \begin{bmatrix}
        Q_{b1}^{\rm SC}-\lambda^{\rm SC}Q_a^{\rm SC}&\frac{1}{2}\left(\mathbf{r}_{b1}^{\rm SC}-\lambda^{\rm SC}\mathbf{r}_{a}^{\rm SC}\right)\\\frac{1}{2}\left(\mathbf{r}_{b1}^{{\rm SC}^{\top}}-\lambda^{\rm SC}\mathbf{r}_{a}^{{\rm SC}^{\top}}\right)&c_{b1}^{\rm SC}-\lambda^{\rm SC}c_a^{\rm SC}
    \end{bmatrix}\geqslant0
\end{equation}

\subsection{Stability Constraints}
The aforementioned state constraint limits the trajectory of active and reactive powers to remain within the defined power safety region. However, when used alone, it does not guarantee system stability as it does not ensure that active and reactive powers will converge to the desired setpoint. Therefore we add the following for stability: 

\emph{Lemma 1:} The grid-connect converter system (\ref{system}) is asymptotically stable with controller (\ref{controller}) if there exists a matrix $\mathbf{P}$ satisfying
\begin{align}
       \mathbf{A}^\top\mathbf{P}+\mathbf{P}\mathbf{A}-&\mathbf{K}^\top\mathbf{B}^\top\mathbf{P}-\mathbf{P}\mathbf{B}\mathbf{K}<0,\label{stability_constraints1}\\
    &\mathbf{P}=\mathbf{P}^\top>0. \label{stability_constraints2}
\end{align}
This condition is found in most standard books on linear system theory (see, e.g.~\cite{hespanha2018linear}).

\subsection{Input Constraints}
Now we look at the input constraints by studying when they would be satisfied when both the state constraint (\ref{stateconstraint}) and the grid voltage (disturbance) constraint (\ref{dconstraint}) are met. This is equivalent to characterizing when the following implication relationship holds: 
\begin{align}\label{ICRelation} 
&\underline{V_{\rm G}}\leqslant V_{\rm G}\leqslant\overline{V_{\rm G}}\;\text{and}\;\frac{P}{\sqrt{P^2+Q^2}}\geqslant \underline{\rm PF}\nonumber\\\Rightarrow &\underline{U}V_{\rm G}\leqslant \|\mathbf{u}\|_2\leqslant \overline{U}V_{\rm G}. 
\end{align} 
We do this through the use of the S-lemma~\cite{Polik2007}. 



1) To facilitate the S-Lemma to parameterize the above implication relationship, we first rewrite the disturbance constraint (\ref{dconstraint}) into the quadratic form by defining $\mathbf{z}=\left[V_{\rm G}^2\;P\;Q\right]^{\top}$:
    \begin{align}\label{dconstraint_quadratic}
    \mathcal{Q}_a^{\rm DC}&=-V_{\rm G}^4+(\underline{V_{\rm G}}^2+\overline{V_{\rm G}}^2)V_{\rm G}^2-\underline{V_{\rm G}}^2\overline{V_{\rm G}}^2\nonumber\\
    &=\mathbf{z}^\top \mathbf{Q}_a^{\rm DC}\mathbf{z}+\mathbf{r}_a^{{\rm DC}^\top}\mathbf{z}+{c}_a^{\rm DC}\geqslant 0
\end{align}
where 
\begin{align*}
   \mathbf{Q}_a^{\rm DC}&=\begin{bmatrix}
    -1 &&\\&0&\\&&0
    \end{bmatrix},\;\mathbf{r}_{a}^{\rm DC}=\begin{bmatrix}
    \underline{V_{\rm G}}^2+\overline{V_{\rm G}}^2&0&0
    \end{bmatrix}^{\top},\nonumber\\
    c_a^{\rm DC}&=-\underline{V_{\rm G}}^2\overline{V_{\rm G}}^2
\geqslant0.
\end{align*}

2) Then, we can also equivalently rewrite the state constraint (\ref{stateconstraint}) in the $\mathbf{z}$-coordinate as follows,
    \begin{align}\label{stateconstraint_quadratic}
    \mathcal{Q}_a^{\rm SC}=\mathbf{z}^\top {\mathbf{Q}}_a^{\rm SC}\mathbf{z}+{\mathbf{r}}_a^{{\rm SC}^\top}\mathbf{z}+{{c}}_a^{\rm SC}\geqslant 0
\end{align}
where
\begin{align*}
   {\mathbf{Q}}_a^{\rm SC}=\begin{bmatrix}
   0&0&0\\ 0&1-\underline{\rm PF}^2 &0\\0&0&-\underline{\rm PF}^2
    \end{bmatrix},\;{\mathbf{r}}_a^{\rm SC}=\begin{bmatrix}
    0&0&0
    \end{bmatrix}^{\top},\; {{c}}_a^{\rm SC}=0.
\end{align*}
such that $\mathcal{Q}_a^{\rm SC}$ and $\mathcal{Q}_a^{\rm DC}$ can be summed up together.

3) Next, we rewrite the input constraint (\ref{inputconstraint}) into quadratic form. By substituting (\ref{controller}) into $\|\mathbf{u}\|_2$, we have
\begin{align}
    \|\mathbf{u}\|_2&=\|-\mathbf{K}(\mathbf{x}-\mathbf{x}_{\rm ref})-\mathbf{B}^{-1}\mathbf{A}\mathbf{x}_{\rm ref}-\mathbf{B}^{-1}\mathbf{E}V_{\rm G}^2\|_2\nonumber\\
    &=\|(\mathbf{K}-\mathbf{B}^{-1}\mathbf{A})\mathbf{x}_{\rm ref}-\mathbf{M}\mathbf{z}\|_2.
\end{align}
where $\mathbf{M}=\left[\mathbf{B}^{-1}\mathbf{E}\;\;\mathbf{K}\right]\in\mathbb{R}^{2\times3}$.
By defining $\underline{\mathbf{U}}=[\underline{U}^2\;0\;0]^{\top}$ and $\overline{\mathbf{U}}=[\overline{U}^2\;0\;0]^{\top}$, the input constraint (\ref{inputconstraint}) can be rewritten as two quadratics:
\begin{subequations}\label{qb}
    \begin{align}
    \mathcal{Q}_{b1}^{\rm IC}=\mathbf{z}^{\top}{\mathbf{Q}_{b1}^{\rm IC}}\mathbf{z}+\mathbf{r}_{b1}^{{\rm IC}^{\top}}\mathbf{z}+{c_{b1}^{\rm IC}}\geqslant0 \label{qb1},\\
    \mathcal{Q}_{b2}^{\rm IC}=\mathbf{z}^{\top}{\mathbf{Q}_{b2}^{\rm IC}}\mathbf{z}+\mathbf{r}_{b2}^{{\rm IC}^{\top}}\mathbf{z}+{c_{b2}^{\rm IC}}\geqslant0, \label{qb2}
\end{align}
\end{subequations}
where 
\begin{align*}
   {\mathbf{Q}_{b1}^{\rm IC}}&=\mathbf{M}^{\top}\mathbf{M},\;{\mathbf{r}_{b1}^{\rm IC}}=-2[(\mathbf{K}-\mathbf{B}^{-1}\mathbf{A})\mathbf{x}_{\rm ref}]^{\top}\mathbf{M}-\underline{\mathbf{U}}^{\top},\\ {c_{b1}^{\rm IC}}&=[(\mathbf{K}-\mathbf{B}^{-1}\mathbf{A})\mathbf{x}_{\rm ref}]^{\top}[(\mathbf{K}-\mathbf{B}^{-1}\mathbf{A})\mathbf{x}_{\rm ref}],\\
   {\mathbf{Q}_{b2}^{\rm IC}}&=-\mathbf{M}^{\top}\mathbf{M},\;{\mathbf{r}_{b2}^{\rm IC}}=2[(\mathbf{K}-\mathbf{B}^{-1}\mathbf{A})\mathbf{x}_{\rm ref}]^{\top}\mathbf{M}+\overline{\mathbf{U}}^{\top},\\ {c_{b2}^{\rm IC}}&=-[(\mathbf{K}-\mathbf{B}^{-1}\mathbf{A})\mathbf{x}_{\rm ref}]^{\top}[(\mathbf{K}-\mathbf{B}^{-1}\mathbf{A})\mathbf{x}_{\rm ref}].
\end{align*}

4) Directly parameterizing (\ref{ICRelation}) using the S-Lemma is challenging, as a single inequality cannot represent the ``and'' conjunction. Therefore, we propose the following lemma to establish a sufficient condition.

\emph{Lemma 2:} Consider quadratics $\mathcal{Q}_a^{\rm DC}\geqslant0$, $\mathcal{Q}_a^{\rm SC}\geqslant0$, and $\mathcal{Q}_{bi}^{\rm IC}\geqslant0$, $i=1,2$. Then, the implication relationship (\ref{ICRelation}) holds, if 
\begin{align}\label{lemma2}
    \mathcal{Q}_a^{\rm DC}+\mathcal{Q}_a^{\rm SC}\geqslant0\Rightarrow \mathcal{Q}_{bi}^{\rm IC}\geqslant 0,
\end{align}

\emph{Proof:}
The relationship (\ref{ICRelation}) can be mathematically restated using quadratics (\ref{dconstraint_quadratic})-(\ref{qb}):
\begin{align}\label{ICRelation_quadratic}
    \mathcal{Q}_a^{\rm DC}\geqslant0\;\text{and}\; \mathcal{Q}_a^{\rm SC}\geqslant0\Rightarrow\mathcal{Q}_{bi}^{\rm IC}\geqslant 0,\;i=1,2.
\end{align}

Since $\mathcal{Q}_a^{\rm DC}\geqslant0\;\text{and}\; \mathcal{Q}_a^{\rm SC}\geqslant0\Rightarrow \mathcal{Q}_a^{\rm DC}+\mathcal{Q}_a^{\rm SC}\geqslant0$ and $\mathcal{Q}_a^{\rm DC}+\mathcal{Q}_a^{\rm SC}\geqslant0\Rightarrow \mathcal{Q}_{bi}^{\rm IC}\geqslant 0$ according to (\ref{lemma2}), it is obvious that (\ref{ICRelation_quadratic}) holds based on chain of implications. \hfill $\Box$

5) We are now prepared to propose the following theorem to parameterize the input constraint.

\emph{Theorem 2:} Given a power setpoint $\mathbf{x}_{\rm ref}$ and corresponding feedback control gain $\mathbf{K}$, if there exist nonnegative constants $\lambda_i^{\rm IC}\geqslant0$ such that
\begin{align}\label{theorem2}
    \mathcal{Q}_{bi}^{\rm IC}-\lambda_i^{\rm IC}\left(\mathcal{Q}_{a}^{\rm SC}
    +\mathcal{Q}_{a}^{\rm DC}\right)\geqslant0,\;\;i=1,2,
\end{align}
then the time-varying nonlinear input constraint (\ref{inputconstraint}) is always satisfied.

\emph{Proof:} According to the S-Lemma, if there exists $\lambda_i^{\rm IC}\geqslant0$ such that (\ref{theorem2}) holds, we have
\begin{align*}
    \mathcal{Q}_a^{\rm DC}+ \mathcal{Q}_a^{\rm SC}\geqslant0\Rightarrow\mathcal{Q}_{bi}^{\rm IC}\geqslant 0,\;i=1,2.
\end{align*}
Then, according to \emph{Lemma 2}, we have (\ref{ICRelation}) holds, which means the input constraint (\ref{inputconstraint}) is satisfied once the disturbance constraint (\ref{dconstraint}) and state constraint (\ref{stateconstraint}) are satisfied. \hfill $\Box$

To eliminate the state variables in (\ref{theorem2}), we convert it into the following LMI:

\begin{align}\label{theorem2_lmi}
    &\begin{bmatrix}
        Q_{bi}^{\rm IC}-\lambda_i^{\rm IC}(Q_a^{\rm SC}+Q_a^{\rm DC})&\!\!\!\!\frac{1}{2}\left(\mathbf{r}_{bi}^{\rm IC}-\lambda_i^{\rm IC}(\mathbf{r}_{a}^{\rm SC}+\mathbf{r}_{a}^{\rm DC})\right)\\\frac{1}{2}\left(\mathbf{r}_{bi}^{{\rm IC}^{\top}}-\lambda_i^{\rm IC}(\mathbf{r}_{a}^{{\rm SC}}+\mathbf{r}_{a}^{{\rm DC}})^{\top}\right)&\!\!\!\!c_{bi}^{\rm IC}-\lambda_i^{\rm IC}(c_a^{\rm SC}+c_a^{\rm DC})
    \end{bmatrix}\nonumber\\&\geqslant0,\;\;i=1,2.
\end{align}

\subsection{Definition of Achievability and Achievable Set}
Based on the state, input and stability constraints parameterized with respect to the control gain $\mathbf{K}$ and power setpoint $\mathbf{x}_{\rm ref}$ in the previous subsections, we are now able to give a series of mathematical definitions about the achievability.

\emph{Definition 1 (Achievability):} A power setpoint $\mathbf{x}_{\rm ref}$ is achievable if there exists a control gain $\mathbf{K}$, such that (\ref{theorem1_lmi}), (\ref{stability_constraints1}), (\ref{stability_constraints2}) and (\ref{theorem2_lmi}) simultaneously hold for some $\mathbf{P}$, $\lambda^{\rm SC}$, $\lambda^{\rm IC}_1$ and $\lambda^{\rm IC}_2$.

\emph{Definition 2 (Achievable Set):} Fixing the control gain $\mathbf{K}$, a set of power setpoints is called an achievable set if each power setpoint $\mathbf{x}_{\rm ref}$ in the set satisfies the conditions for achievability with this $\mathbf{K}$.

Definitions 1 and 2 serve distinct but complementary purposes. Definition 1 can be used to assess the achievability of all power setpoints when the control gain is allowed to be adjusted dynamically in real time. In contrast, Definition 2 identifies the set of power setpoints that are achievable under a fixed control gain, which is particularly useful in scenarios where the controller parameters cannot be updated on demand.


\section{Controller Optimization and Achievability Analysis}\label{C4}
In the previous section, we parameterized the state, input, and stability constraints in terms of power setpoints and control gain. In this section, we will explore how these parameterized constraints can be employed to optimize the safe controller and maximize the achievability set.
\subsection{Controller Ensuring State, Input and Stability Contraints}
For an arbitrarily given power setpoint $\mathbf{x}_{\rm ref}$, we aim to find a feedback control gain $\mathbf{K}$ such that (\ref{theorem1_lmi}), (\ref{stability_constraints1}), (\ref{stability_constraints2}) and (\ref{theorem2_lmi}) simultaneously hold. This can be formulated as the following optimization problem:

\begin{align*}
    {\rm Problem \;P1}:\;\min_{\mathbf{K},\mathbf{P},\lambda^{\rm SC},\lambda^{\rm IC}_1,\lambda^{\rm IC}_2} \;&\|\mathbf{K}\|_{\rm F}^2\\
    {\rm s.t.}\; &\text{BMI} \;(\ref{stability_constraints1})\nonumber\\
    &\text{LMIs} \;(\ref{theorem1_lmi}),(\ref{stability_constraints2})\;\text{and}\;(\ref{theorem2_lmi})\nonumber
\end{align*}
where $\|\cdot\|_{\rm F}$ denotes the Frobenius Norm of a matrix. It should be noted that constraint (\ref{stability_constraints1}) is a bilinear matrix inequality (BMI) when simultaneously searching for both $\mathbf{P}$ and $\mathbf{K}$ in Problem P1. This problem can be approached using nonlinear programming algorithms, including methods like the active-set method, sequential quadratic programming (SQP), or interior-point methods. The solution to Problem P1 provides a feasible controller (\ref{controller}) that stabilizes the active and reactive power of the inverter system (\ref{system_pq}) to the desired setpoint $\mathbf{x}_{\rm ref}$, while ensuring compliance with input constraints, state constraints, and closed-loop stability. 

Problem P1 provides the largest achievable set because each setpoint can be assigned a tailored control gain. However, such a set is not practically applicable due to significant limitations: First, this approach requires substantial computer memory for the local controller to store the control gain/setpoint pairs, increasing hardware costs and introducing a computational burden associated with searching for the appropriate control gain in the library. Second, it can only check a finite number of setpoints, but $\xref$ is likely to take on continuous values in practice. Both problems will be solved in the following subsections. 

\subsection{Maximizing the Achievable Set for a Single Control Gain}

To overcome the first challenge, we can select a stabilizing control gain $\bd{K}$. 
Then, denoting $\mathbf{\Lambda}=\left[\lambda^{\rm SC}\;\lambda^{\rm IC}_1\;\lambda^{\rm IC}_2\right]^{\top}$, we check the achievability of each power setpoint by solving the following Problem P2. 
\begin{align*}
    {\rm Problem \;P2}:\;\min_{\lambda^{\rm SC},\lambda^{\rm IC}_1,\lambda^{\rm IC}_2} \;&\|\mathbf{\Lambda}\|_2^2\\
    {\rm s.t.}\; &\text{LMIs} \;(\ref{theorem1_lmi})\;\text{and}\;(\ref{theorem2_lmi})\nonumber
\end{align*}
This is now a Semidefinite Programming (SDP) problem and can be solved using a number of solvers~\cite{Boyd2004}. 

There are many $\bd{K}$ that will satisfy the constraints in P2. Our goal is to pick the one that results in the largest achievable set. But there is no easy formula for computing the area of the set. Therefore, we take a statistical approach, as proposed in \cite{Ma2024control}. This approach estimates the area by sampling uniformly at random within a predefined domain. We introduce a performance metric, termed the achievability rate, which is defined as follows:
\begin{align} f(\mathbf{K}) = \frac{\text{Number of achievable setpoints} }{\text{Total number of sampled setpoints} }. \end{align}

\begin{figure}[t!]
		\centering
		\includegraphics[width=0.88\columnwidth]{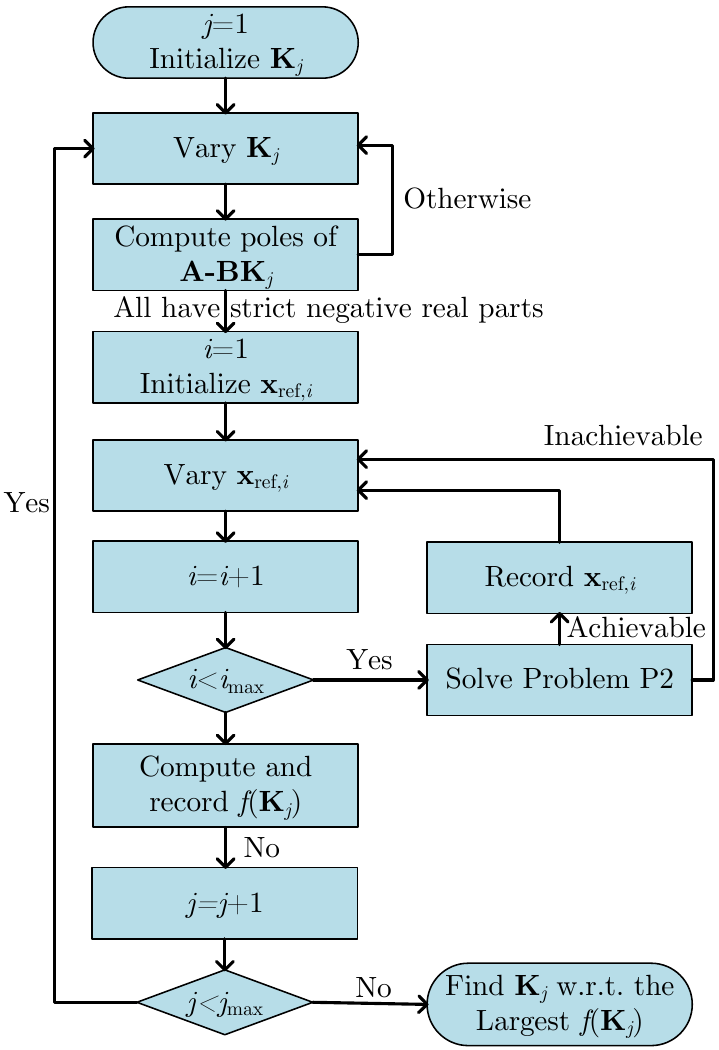}
		\caption{Flowchart illustrating the Monte Carlo simulation for maximizing the achievable set by optimizing over the control gain $\mathbf{K}$. Here, $i_{\rm max}$ and $j_{\rm max}$ represent the total samples of $\mathbf{x}_{\rm ref}$ and $\mathbf{K}$, respectively.}
		\label{algorithm}
						\vspace{-1 em}
	\end{figure}

This formulation enables the optimization of $f(\mathbf{K})$ by adjusting $\mathbf{K}$ via simulation-based optimization techniques. Importantly, since the control gain matrix $\mathbf{K}$ consists of only four parameters, a direct Monte Carlo simulation is computationally efficient for this purpose. Specifically, the optimization involves systematically exploring the parameter space of $\mathbf{K}$ by sweeping through the values of its elements. The detailed procedure for this optimization process is illustrated in Fig. \ref{algorithm}.

\subsection{Extending Achievable Set to Continous Region}
To address the second challenge regarding the discreteness of the achievable set, we employ the convex hull method to extend the isolated achievable points to a continuous region.  By fixing $\mathbf{K}$, $\mathbf{P}$, and $\mathbf{\Lambda}$, all the constraints (\ref{theorem1_lmi}), (\ref{stability_constraints1}), (\ref{stability_constraints2}) and (\ref{theorem2_lmi}) become linear with respect to the setpoint $\mathbf{x}_{\rm ref}$. Based on this, we propose the following lemma:

\emph{Lemma 3:} Any setpoint within the triangular region formed by three achievable setpoints under the same parameters $\mathbf{K}$, $\mathbf{P}$, $\lambda^{\rm SC}$, $\lambda_1^{\rm IC}$ and $\lambda_2^{\rm IC}$, is also achievable.

 \emph{Proof:} Since the constraints are linear with respect to $\mathbf{x}_{\rm ref}$, they can be expressed in the form:
 \begin{equation}\label{convexhull_1}
     \mathbf{H}\mathbf{x}_{\rm ref}\geqslant \mathbf{b}
 \end{equation}
 where $\mathbf{H}$ and $\mathbf{b}$ are derived from (\ref{theorem1_lmi}), (\ref{stability_constraints1}), (\ref{stability_constraints2}) and (\ref{theorem2_lmi}). 
 
 Let $\mathbf{x}_{{\rm ref}i}\in \mathbb{R}^2$, $i=1,2,3$, be three power setpoints satisfying the constraints:
  \begin{equation}\label{convexhull_2}
     \mathbf{H}\mathbf{x}_{{\rm ref}i}\geqslant \mathbf{b}, \; i=1,2,3.
 \end{equation}
Any setpoint $\mathbf{x}_{\rm ref}$ within the convex hull of $\mathbf{x}_{{\rm ref}1},\mathbf{x}_{{\rm ref}2}$, and $\mathbf{x}_{{\rm ref}3}$, can be defined as
 \begin{align}\label{convexhull_3}
     \mathbf{x}_{\rm ref}=\xi_1\mathbf{x}_{{\rm ref}1}+\xi_2\mathbf{x}_{{\rm ref}2}+\xi_3\mathbf{x}_{{\rm ref}3}
 \end{align}
 where $\xi_1,\xi_2,\xi_3\geqslant0$, $\xi_1+\xi_2+\xi_3=1$.

Substituting (\ref{convexhull_3}) into $\mathbf{H}\mathbf{x}_{\rm ref}$, and using (\ref{convexhull_2}), we obtain
 \begin{align}
     \mathbf{H}\mathbf{x}_{\rm ref}=&\xi_1\mathbf{H}\mathbf{x}_{{\rm ref}1}+\xi_2\mathbf{H}\mathbf{x}_{{\rm ref}2}+\mathbf{H}\xi_3\mathbf{x}_{{\rm ref}3}\nonumber\\
\geqslant&\xi_1\mathbf{b}+\xi_2\mathbf{b}+\xi_3\mathbf{b}=\mathbf{b}.
 \end{align}
 
According to \emph{Definition 1},  this proves the achievablity of any power setpoint $\mathbf{x}_{\rm ref}$ within the convex hull of three achievable setpoints, $\mathbf{x}_{{\rm ref}1}$, $\mathbf{x}_{{\rm ref}2}$ and $\mathbf{x}_{{\rm ref}3}$. \hfill $\Box$

\begin{figure}[t!]
		\centering
		\includegraphics[width=0.88\columnwidth]{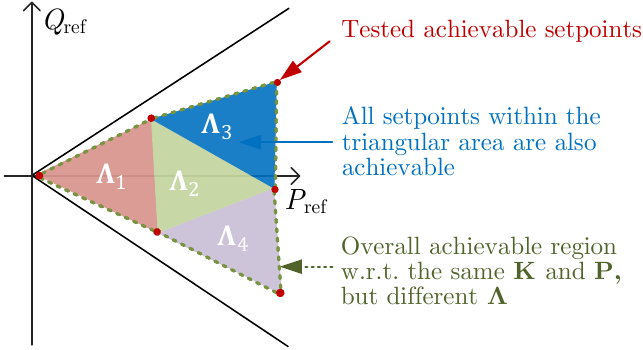}
		\caption{Illustration of applying Lemma 3 to identify continuous achievable regions. The achievability of a finite number of setpoint samples is verified by checking whether the constraints (\ref{theorem1_lmi}), (\ref{stability_constraints1}), (\ref{stability_constraints2}), and (\ref{theorem2_lmi}) are satisfied for a specific set of $\mathbf{K}$, $\mathbf{P}$, and $\mathbf{\Lambda}$. Setpoints not explicitly simulated but located within the triangular area (depicted by the colored triangles) are also deemed achievable. This approach leverages a discrete simulation method to derive a continuous achievable region.}
		\label{convexhullfig}
						\vspace{-1 em}
	\end{figure}

As visually demonstrated in Fig. \ref{convexhullfig}, \emph{Lemma 3} provides a means to determine continuously achievable regions.

\section{Case studies}\label{C5}
In this section, we evaluate the performance of the proposed state-and-input-constrained power controller. The test system consists of a two-bus grid-connected inverter, as illustrated in Fig. \ref{systemfig}. The system parameters are listed in Table \ref{parameters}. The controller is implemented as described in (\ref{controller}). The grid voltage, $V_{\rm G}$, is assumed to be measurable but is designed to vary randomly within the range of $\underline{V_{\rm G}}$ to $\overline{V_{\rm G}}$. The power factor limit in the state constraint $\underline{\rm PF}=0.95$.

\begin{table}[t!]
\setlength{\tabcolsep}{5pt}
\centering
\caption{Parameter setting of grid-connected IBR}
\renewcommand\arraystretch{1.5}
\label{parameters}
\begin{tabular}{p{1cm}<{\raggedright}p{2.5cm}<{\raggedright}p{1cm}<{\raggedright}p{1cm}<{\raggedright}}
\hline\hline
 Par. & Value  & Par. & Value \\ \hline
 $R$ & $0.12 \;\Omega$ & $L$ & $4$ mH\\
 $\underline{V_{\rm G}}$ & $105.6$ V & $\overline{V_{\rm G}}$ & $114.4$ V\\
  $\underline{U}$ & $104.5$ V & $\overline{U}$ & $115.5$ V\\
  $\omega$ & $314$ rad/s & $f$ & $50$ Hz
	\\ \hline \hline
		\end{tabular}
				\vspace{-1 em}
	\end{table}

We conduct several case studies to validate the proposed approach. First, we demonstrate the procedure for determining a static state feedback control gain $\mathbf{K}$ and the associated continuous achievable region. Next, we evaluate the method's capability to maintain operation during fault conditions. Finally, we compare its control performance against conventional approaches, including LQR and MPC. All simulations and analyses are performed using MATLAB.

\subsection{Optimization of Control Gain and Achievable Region}
\begin{figure}[t!]
		\centering
		\includegraphics[width=0.85\columnwidth]{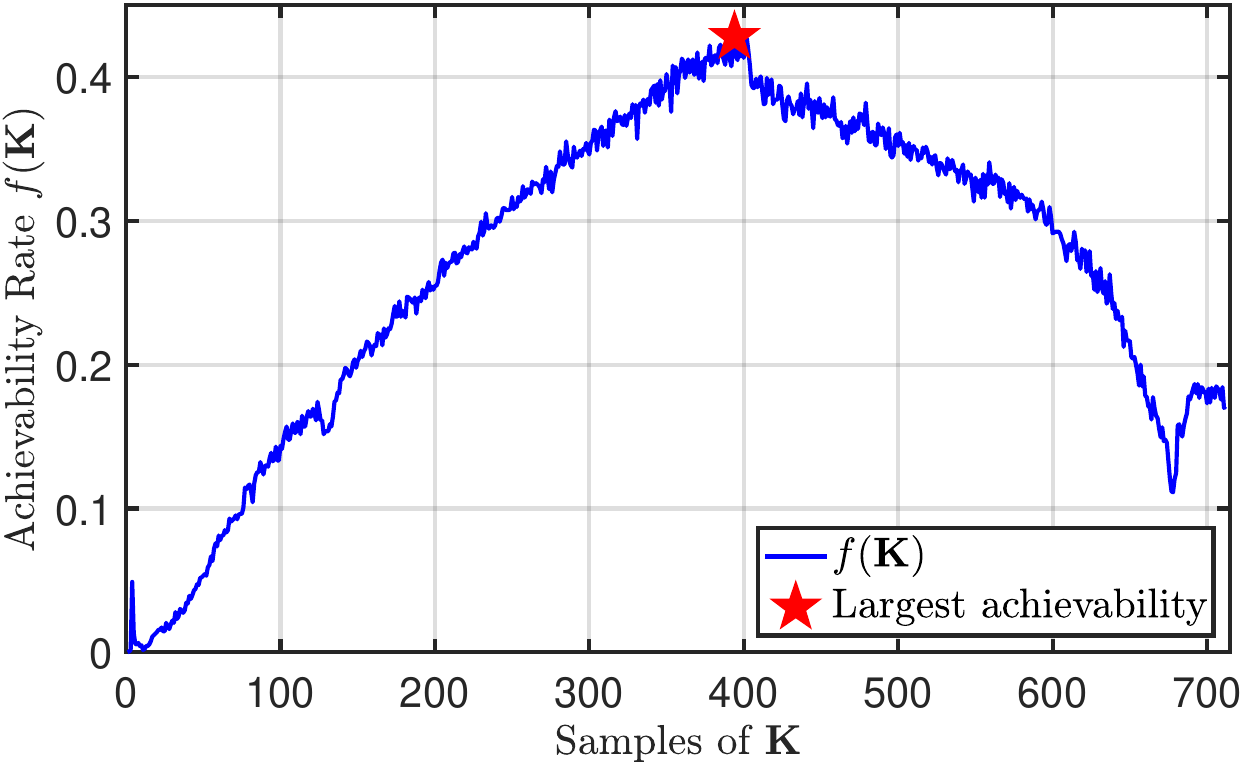}
		\caption{Optimization results for the achievability rates, $f(\mathbf{K})$, with respect to the $k_{11}$ component of $\mathbf{K}$ in the Monte Carlo simulation. Out of 1000 tested samples, 719 satisfy the stability constraints. The achievability rate $f(\mathbf{K})$ reaches its maximum value when $k_{11} = -0.0015$.}
		\label{achievable_rate}
	\end{figure}

Using the proposed algorithm shown in Fig. \ref{algorithm}, the achievable region can be maximized by optimizing the gain matrix $\mathbf{K}$. To achieve this, the four elements of $\mathbf{K}$, $(k_{11}, k_{12}, k_{21}, k_{22})$, are uniformly sampled within a neighborhood around zero. The optimization problem P2 is then iteratively solved for each sample in a Monte Carlo simulation. Fig. \ref{achievable_rate} illustrates the optimization results of the achievability rate $f(\mathbf{K})$ as a function of the first element $k_{11}$ of $\mathbf{K}$. The optimized gain matrix is determined to be ${\mathbf{K}=\begin{bmatrix}
    -0.0015&-0.0003\\-0.4028&-0.3211
\end{bmatrix}}$.

\begin{figure}[t!]
		\centering
		\includegraphics[width=0.85\columnwidth]{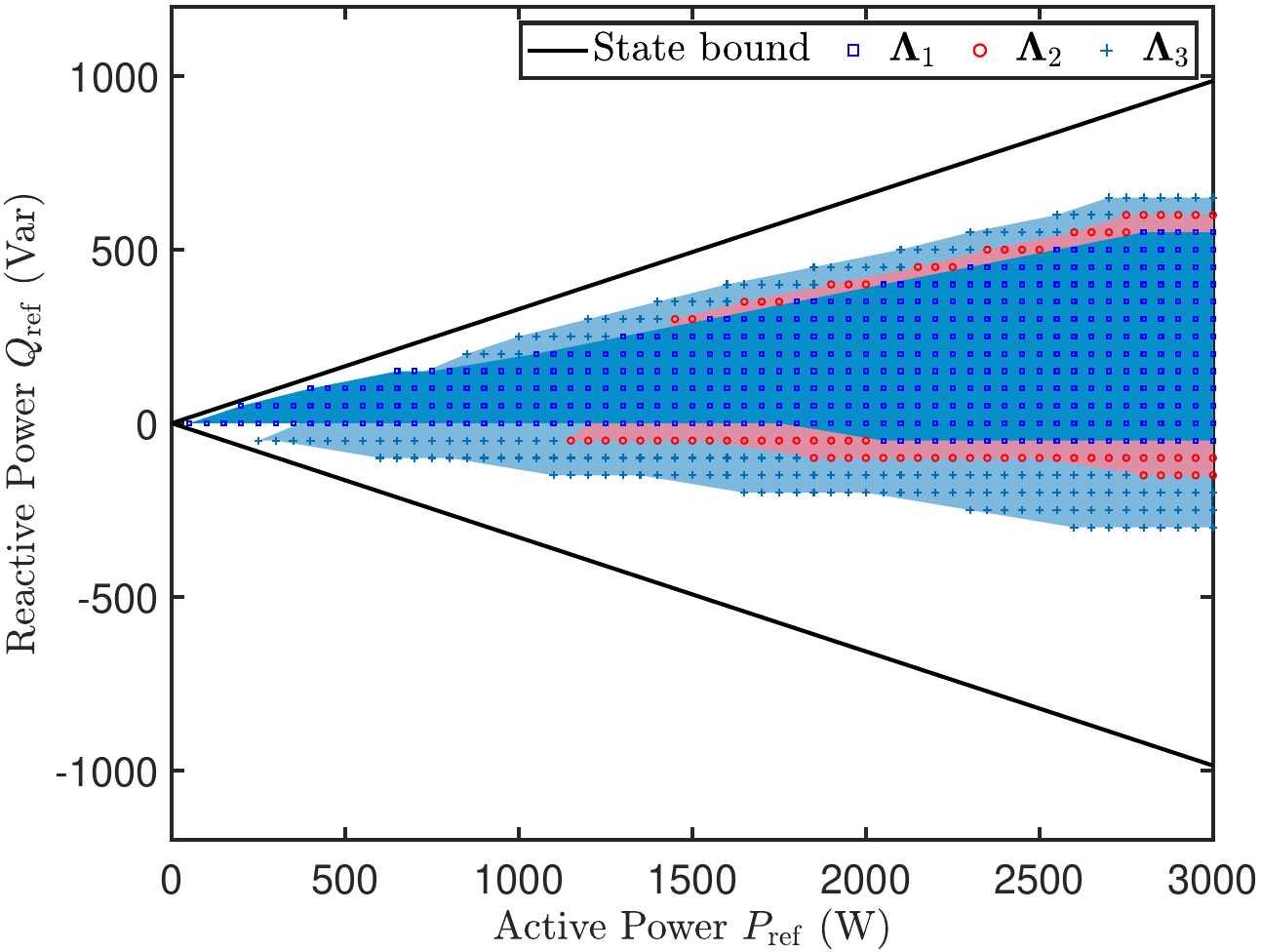}
		\caption{The achievable regions corresponding to the optimal $\mathbf{K}$ are illustrated by the three colored areas, each determined for a fixed value of $\mathbf{\Lambda}$. According to \emph{Lemma 3}, all power setpoints within these colored regions, defined by the achievable samples, are also guaranteed to be achievable. The union of the three areas represents the final continuous achievable region associated with the optimal $\mathbf{K}$. }
		\label{AS_diff_lambda}
	\end{figure}
    
To evaluate the achievable region, the power setpoints $P_{\rm ref}$ and $Q_{\rm ref}$ are sampled uniformly within specified ranges: $P_{\rm ref} \in [0, 3000]$ W and $Q_{\rm ref} \in [-1000, 1000]$ Var. These sampled setpoints are represented as dots in Fig. \ref{AS_diff_lambda}. Based on \emph{Lemma 3}, the achievable region can be determined using these discrete samples, provided the setpoints are achievable under a consistent set of parameters, including $\mathbf{K}$, $\mathbf{P}$, and $\mathbf{\Lambda}$. In Fig. \ref{AS_diff_lambda}, the optimal gain matrix $\mathbf{K}$ is fixed, and the achievability of the sampled setpoints is evaluated for three different values of $\mathbf{\Lambda}$. The continuous achievable regions corresponding to each $\mathbf{\Lambda}$ are depicted as colored areas in the figure. The union of these areas is the achievable region of $\mathbf{K}$.

\begin{figure}[t!]
		\centering
		\includegraphics[width=0.95\columnwidth]{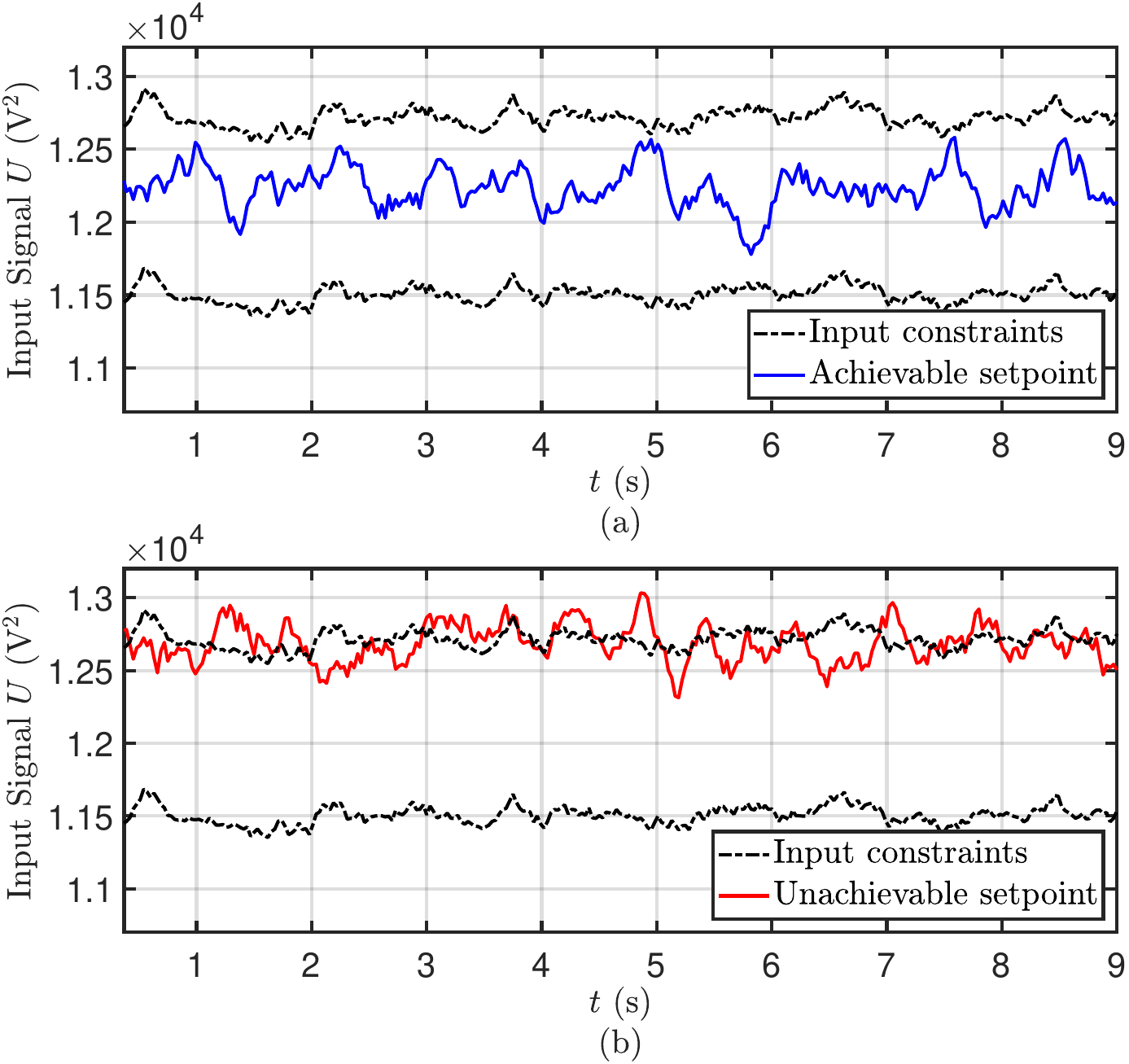}
		\caption{The input signal of the proposed control method. Fig. (a) shows that when the setpoint is selected within the achievable region ($\left[P_{\rm ref}\;\;Q_{\rm ref}\right]^{\top}=\left[1300 \;\text{W}\;\;120 \;\text{Var}\right]^{\top}$), the control signal remains consistently bounded within the input constraints. In contrast, Fig. (b) demonstrates that when the setpoint is outside the achievable region ($\left[P_{\rm ref}\;\;Q_{\rm ref}\right]^{\top}=\left[1800 \;\text{W}\;\;550 \;\text{Var}\right]^{\top}$), the control signal occasionally exceeds the input constraints.}
		\label{input_inside_outside}
	\end{figure}

\begin{figure}[tb!]
		\centering
		\includegraphics[width=0.95\columnwidth]{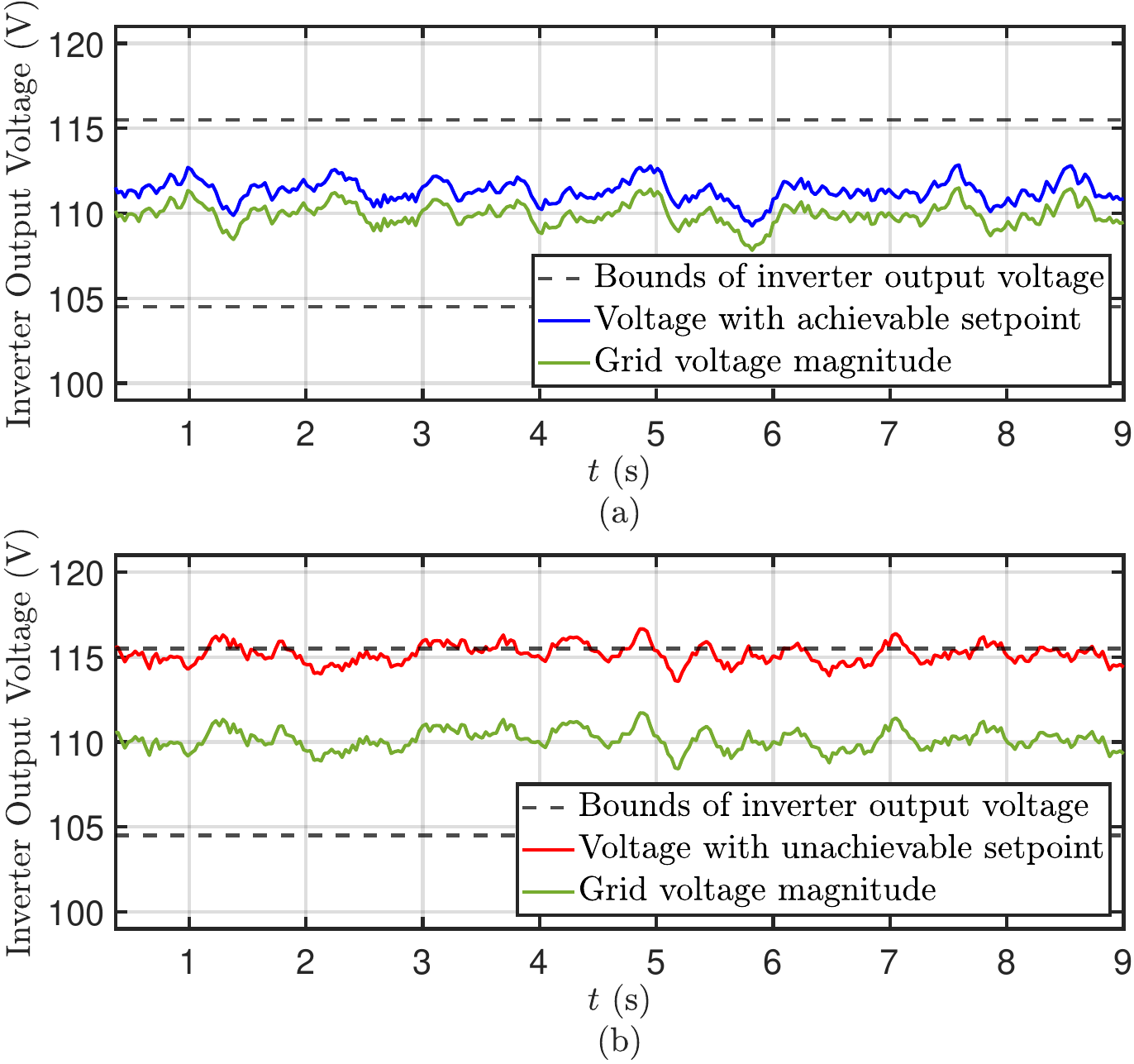}
		\caption{The inverter output voltage of the proposed control method. Fig. (a) shows that when the setpoint is selected within the achievable region ($\left[P_{\rm ref}\;\;Q_{\rm ref}\right]^{\top}=\left[1300 \;\text{W}\;\;120 \;\text{Var}\right]^{\top}$), the inverter output voltage is bounded within the voltage constraint. Fig. (b) illustrates that when the setpoint is selected outside ($\left[P_{\rm ref}\;\;Q_{\rm ref}\right]^{\top}=\left[1800 \;\text{W}\;\;550 \;\text{Var}\right]^{\top}$), the inverter output voltage sometimes violates the voltage constraints.}
		\label{voltage_inside_outside}
	\end{figure}

To illustrate the practical application of the determined achievable region, we compare the control input signals for power setpoints located inside and outside the achievable region. In Fig. \ref{input_inside_outside}, the control input satisfies the input constraints when the setpoints are chosen within the achievable region. In contrast, it violates the input constraints when the setpoints are outside the achievable region. Additionally, Fig. \ref{voltage_inside_outside} presents the inverter output voltages corresponding to the two control signals under the respective setpoints, highlighting the impact of achievable and non-achievable setpoints on system performance.

\begin{figure}[t!]
		\centering
		\includegraphics[width=0.95\columnwidth]{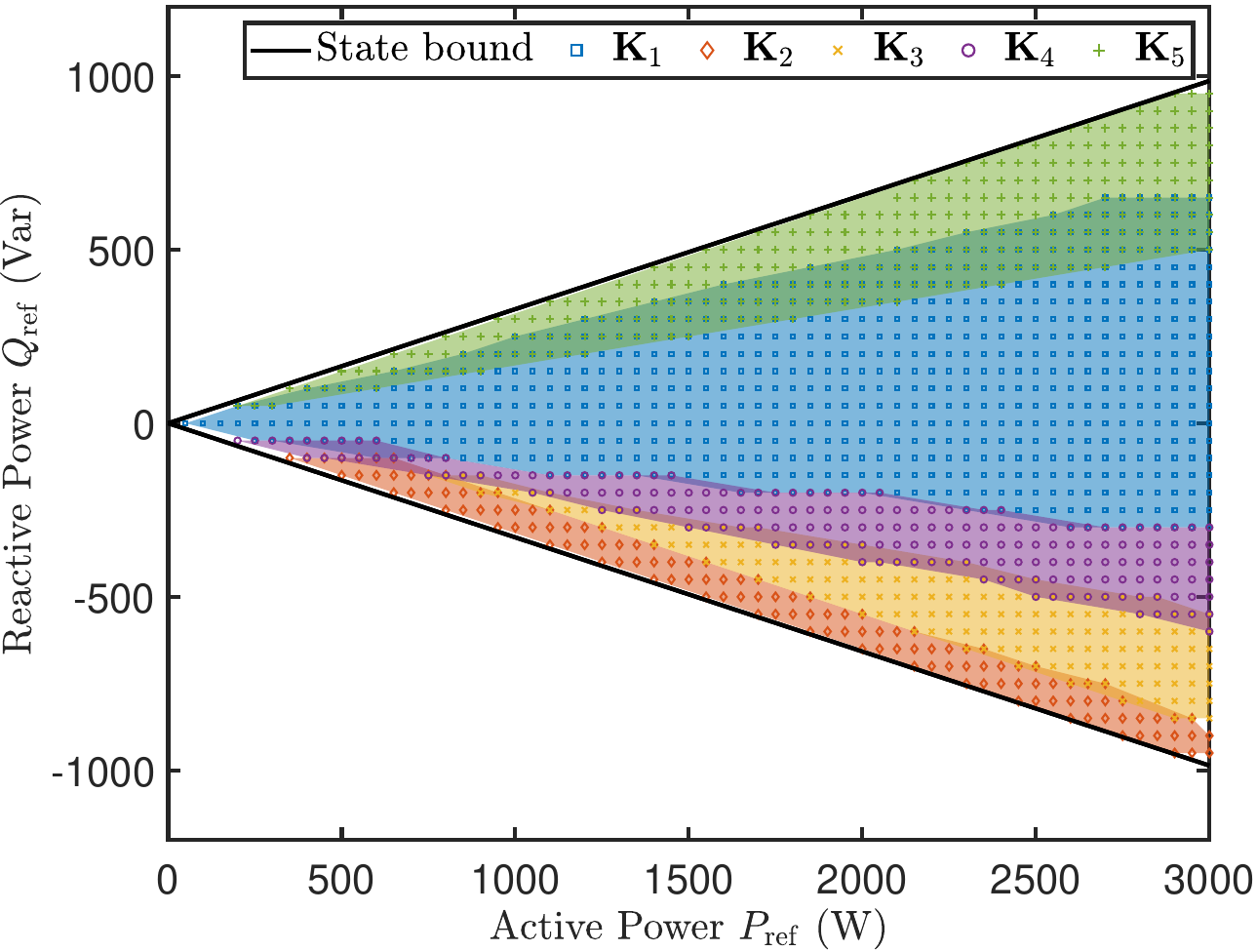}
		\caption{The achievable regions corresponding to five different $\mathbf{K}$ values are represented by the five colored areas in the figure. As shown, all power setpoints within the state-constrained region are achievable. By storing only these five $\mathbf{K}$ values and switching between them as needed when changing the power setpoint, the inverter output power can be regulated to any desired setpoint within the region. }
		\label{AS_different_K}
	\end{figure}

We also examine the achievability of regions that are not achievable with the optimal $\mathbf{K}$ in Fig. \ref{AS_diff_lambda}. By solving P1, we determine that all power setpoints within the state-constrained area are indeed achievable under different configurations of $\mathbf{K}$. This raises an important question: can a finite set of $\mathbf{K}$ values be identified to cover the entire region? Such a solution would enable efficient memory usage by requiring only a change in $\mathbf{K}$ when adjusting the power setpoint. As shown in Fig. \ref{AS_different_K}, the entire region, within the scope of the tested $P_{\rm ref}$ and $Q_{\rm ref}$, can be fully covered using just five different $\mathbf{K}$ values.

\subsection{Ride-Through Capability Test}
\begin{figure}[t!]
		\centering
		\includegraphics[width=0.9\columnwidth]{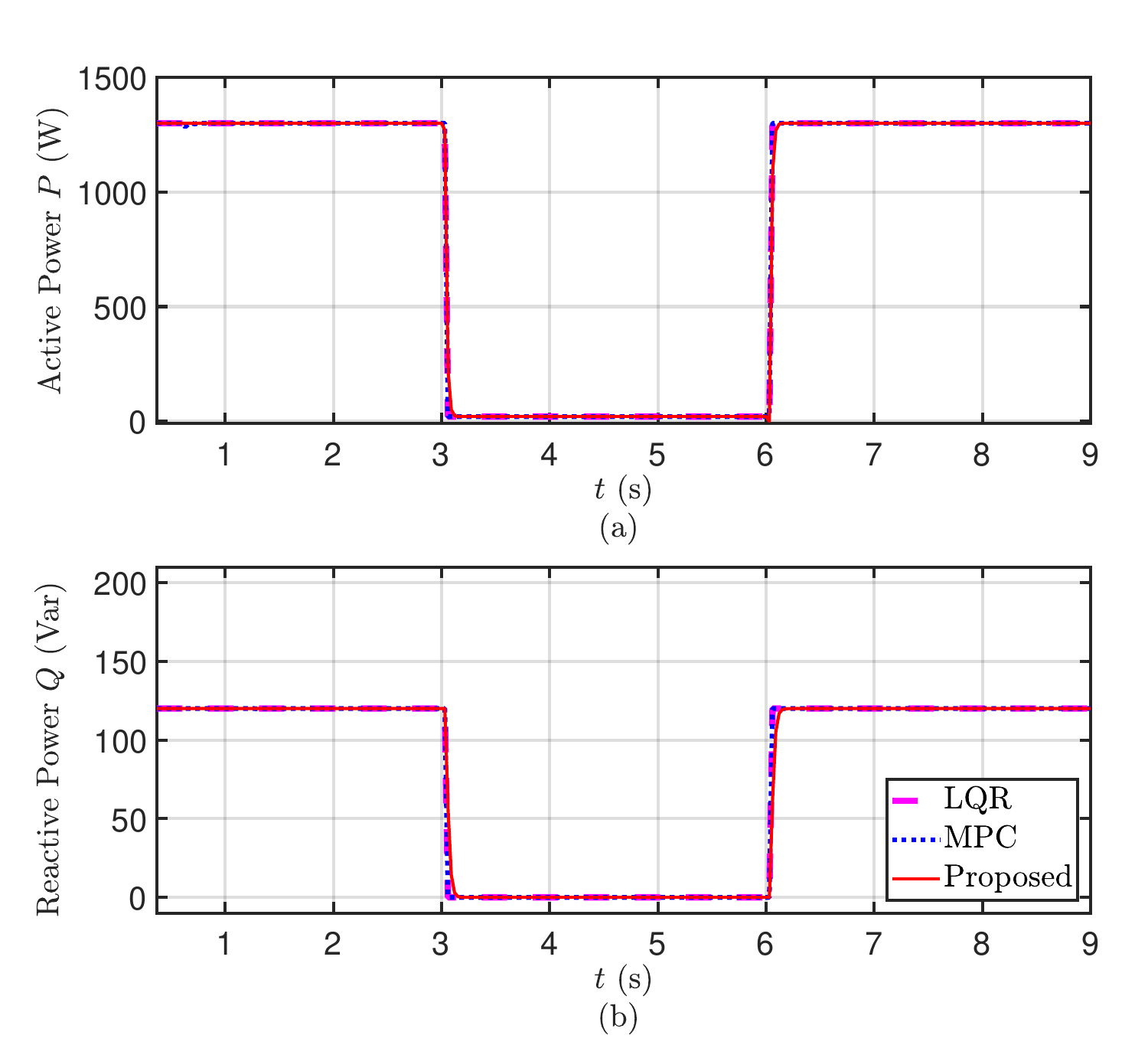}
		\caption{Comparison of active and reactive power regulation results using the proposed method, LQR, and MPC. From 3 seconds to 6 seconds, a fault scenario is simulated by reducing the power setpoint to $[P_{\rm ref}\;\; Q_{\rm ref}]^{\top} = [20\;\text{W}\;\; 0\;\text{Var}]^{\top}$ to facilitate a ride-through event. The active power is not set to zero to prevent an undefined power factor.}
		\label{state_ride_through}
	\end{figure}

\begin{figure}[t!]
		\centering
		\includegraphics[width=0.9\columnwidth]{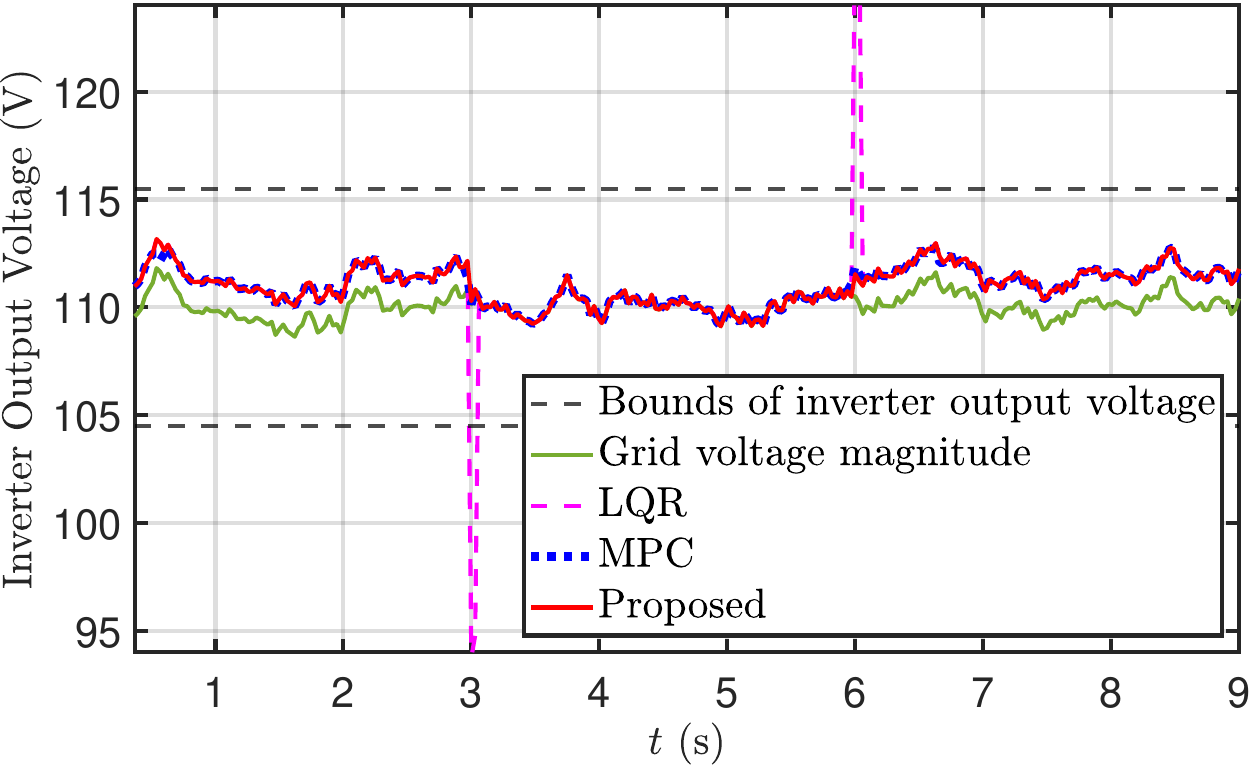}
		\caption{Comparison of inverter output voltages using the proposed method, LQR, and MPC. All three methods successfully maintain the inverter output voltage in alignment with the grid voltage during the ride-through event. The results show that both the proposed method and MPC consistently satisfy the voltage constraints, while LQR fails to meet the constraints after a sudden change in the power setpoint. }
		\label{Inverter_outputvoltage_compare.pdf}
	\end{figure}

\begin{figure}[t!]
		\centering
		\includegraphics[width=0.9\columnwidth]{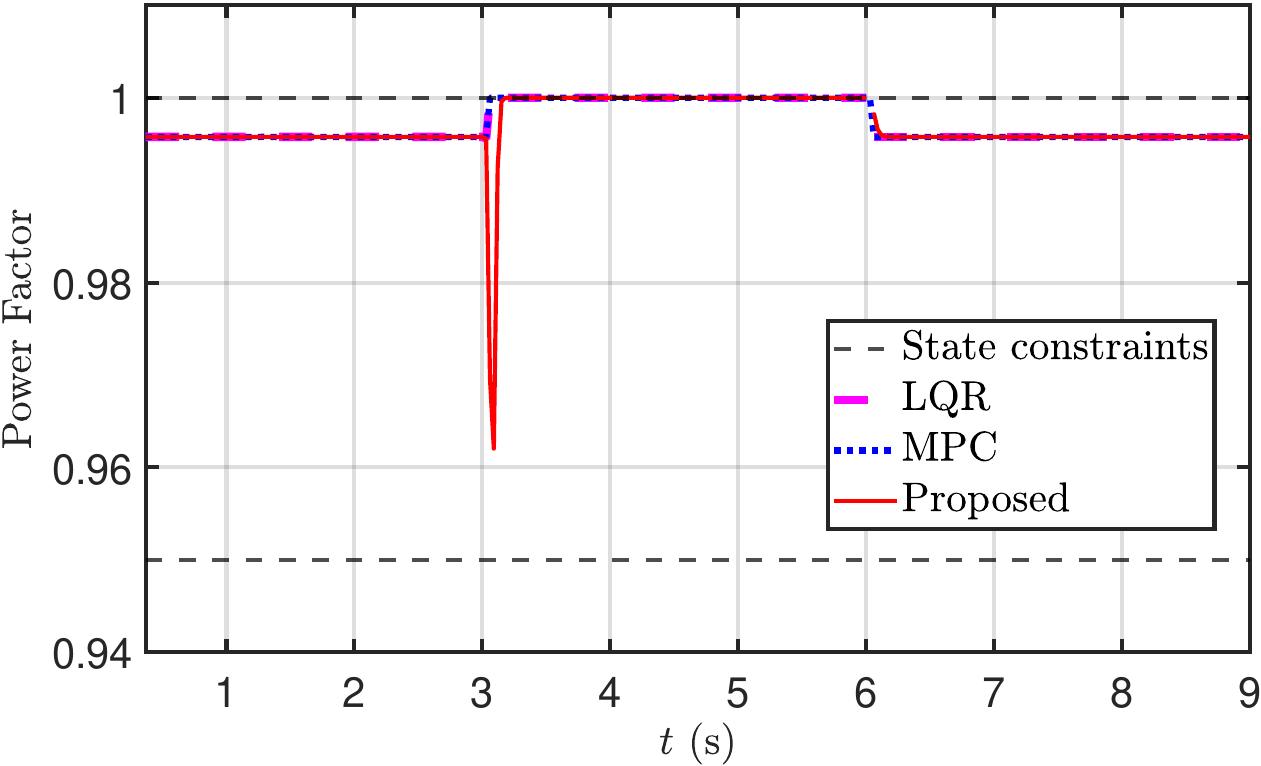}
		\caption{Comparison of power factors using the proposed method, LQR, and MPC. All three methods satisfy the state constraints.}
		\label{PF_compare}
	\end{figure}

\begin{figure}[t!]
		\centering
		\includegraphics[width=0.9\columnwidth]{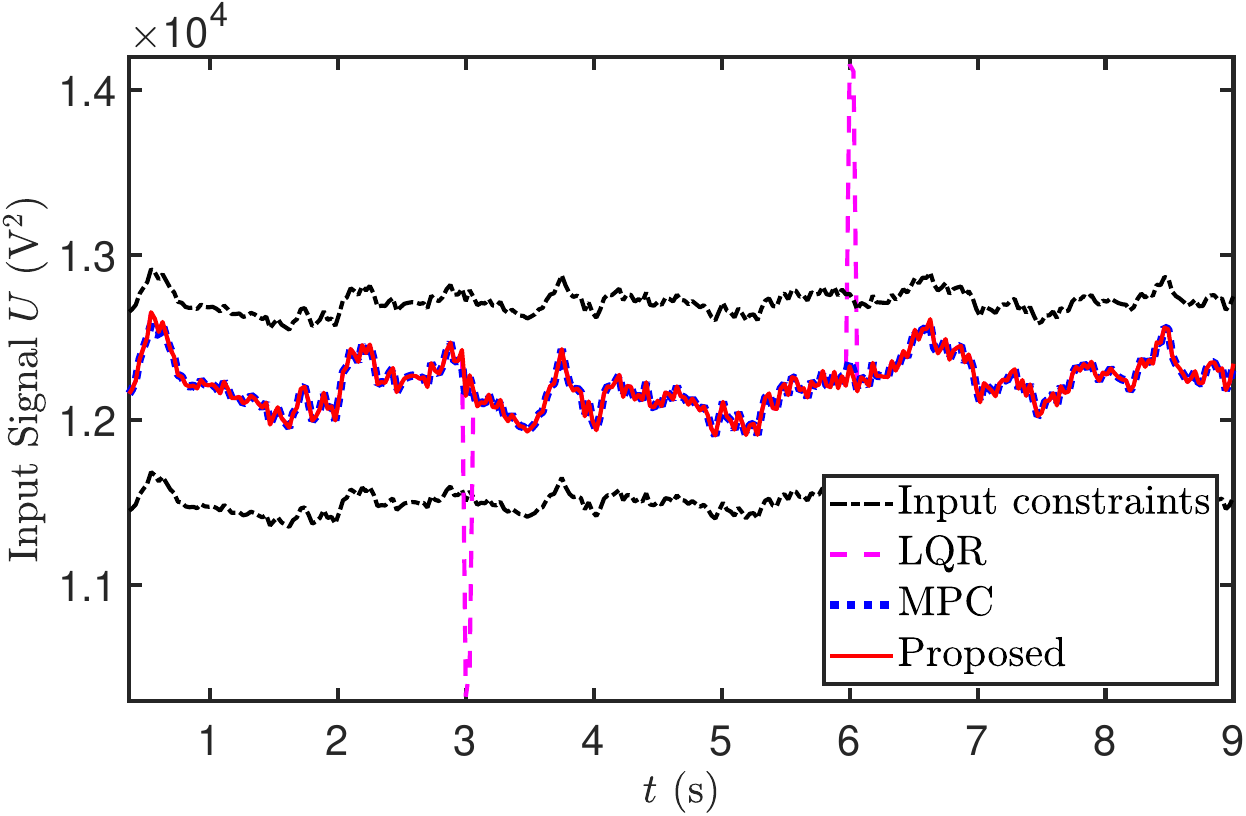}
		\caption{Comparison of input signals for the proposed method, LQR, and MPC. The results indicate that both the proposed method and MPC consistently satisfy the time-varying input constraints, whereas LQR fails to meet these constraints following a sudden change in the power setpoint.}
		\label{input_ride_through}
	\end{figure}

In this subsection, we evaluate the proposed controller's ability to accurately track power setpoints while ensuring stability and satisfying state and input constraints during a ride-through event. In power systems, a ride-through event refers to a scenario where a system, such as a wind turbine or inverter, remains operational during a grid disturbance (e.g., a fault). Instead of shutting down, the system adjusts its operation, such as reducing the power setpoint to zero, while maintaining its output voltage. To simulate this scenario, we change the setpoint from normal operation at $[P_{\rm ref}\;\; Q_{\rm ref}]^{\top} = [1300\;\text{W}\;\; 120\;\text{Var}]^{\top}$ to $[P_{\rm ref}\;\; Q_{\rm ref}]^{\top} = [20\;\text{W}\;\; 0\;\text{Var}]^{\top}$ during the fault period from 3 seconds to 6 seconds. Note that $P_{\rm ref}$ is not set to zero to avoid an undefined power factor (0/0). Fig. \ref{state_ride_through} demonstrates that the proposed controller accurately regulates the active and reactive powers to their desired setpoints. Fig. \ref{Inverter_outputvoltage_compare.pdf} shows that the inverter output voltage remains consistent with the grid voltage during the fault. Additionally, Figs. \ref{PF_compare}–\ref{input_ride_through} validate the effectiveness of the proposed method in ensuring the satisfaction of state and input constraints during the ride-through event.

\subsection{Comparison of Proposed Controller Performance with Traditional Methods}
We compare the control performance of the proposed method with two traditional methods: LQR and MPC. Since the future values of $V_{\rm G}$ are unknown, we use $\underline{U}\overline{V_{\rm G}}$ and $\overline{U}\underline{V_{\rm G}}$ as the lower and upper constraint bounds, respectively, during the prediction stage of MPC to ensure the satisfaction of (\ref{inputconstraint}) at all times. The results, shown in Figs. \ref{state_ride_through}–\ref{input_ride_through}, evaluate power control performance and the ability to enforce state and input constraints.

The figures demonstrate that both the proposed method and MPC reliably satisfy the constraints, whereas LQR fails to meet the input constraint during the transition of power setpoints due to its inability to strictly enforce constraints. While MPC exhibits a slightly faster dynamic response following setpoint changes, its computational time for generating a single control signal at each time step is significantly higher, averaging $0.023$ seconds compared to $5 \times 10^{-5}$ seconds for both LQR and the proposed method. This highlights the superior computational efficiency of the proposed approach.

\section{Conclusion}\label{C6}
This paper presents a novel control framework for grid-connected inverter-based resources (IBRs) that addresses the challenges of simultaneous constraint satisfaction, computational efficiency, and dynamic grid conditions. By leveraging a static state-feedback controller, the proposed method ensures stability while adhering to both state and input constraints. The introduction of the achievability concept provides a systematic way to evaluate whether power setpoints can be reliably tracked under time-varying and nonlinear constraints. We use S-lemma to parameterize and maximize the achievable region. 

The proposed approach significantly reduces online computational complexity by precomputing a finite set of control gains, allowing system operators to seamlessly switch between them as needed. Simulation results validate the effectiveness of the framework in maintaining reliable performance under time-varying disturbances and ensuring the operational safety of IBRs. Future work will focus on applying the proposed method to system-level studies, such as black start processes, where the achievable region can be utilized as a static constraint to enhance safety, stability, and computational efficiency during operations involving only IBRs.
\bibliographystyle{IEEEtran}
	\bibliography{reference}
\end{document}